\documentclass[12pt]{article}
\usepackage{feynmf}
\usepackage{latexsym,bbm}
\usepackage{amsmath,amssymb}

\newcommand{\e}{\mathrm{e}}
\newcommand{\FMslash}[1]{#1\mspace{-8.0mu}/}
\newcommand{\veva}{\ensuremath{\langle A\rangle_\beta}} 
\newcommand{\qslash}{\FMslash{q}}
\newcommand{\kslash}{\FMslash{k}}
\newcommand{\pslash}{\FMslash{p}}
\newcommand{\dslash}{\FMslash{\partial\mspace{-1.0mu}}\mspace{1.0mu}}
\newcommand{\nb}[1]{\ensuremath{n_\mathrm{B}(#1)}} 
\newcommand{\nf}[1]{\ensuremath{n_\mathrm{F}(#1)}} 
\newcommand{\qvec}{\ensuremath{\vec{\mspace{1.0mu}q}}}
\newcommand{\pvec}{\ensuremath{\vec{\mspace{1.0mu}p}\mspace{1.0mu}}}
\newcommand{\intdrei}[1]
    {\ensuremath{\int\!\!\frac{\mathrm{d}^3 #1}{(2\pi)^3}\,}}
\newcommand{\intvier}[1]
    {\ensuremath{\int\!\!\frac{\mathrm{d}^4 #1}{(2\pi)^4}\,}}
\newcommand{\intvierx}{\ensuremath{\int\! \mathrm{d}^4 x}\,}
\newcommand{\intdreix}{\ensuremath{\int\! \mathrm{d}^3 x}\,}
\newcommand{\deltav}{\ensuremath{\delta^{(4)}}}
\newcommand{\shortoverline}[1]
    {\mspace{1mu}\overline{\mspace{-1mu}#1\mspace{-1mu}}\mspace{1mu}}
\newcommand{\kaisoverline}[3]
    {\mspace{#1}\overline{\mspace{-#1}#3\mspace{-#2}}\mspace{#2}}
\newcommand{\Psibar}{\kaisoverline{1.0mu}{2.0mu}\Psi}
\newcommand{\Jbar}{\kaisoverline{3mu}{0mu}{J}}
\DeclareMathOperator{\tr}{tr}

\DeclareMathOperator{\sign}{sign}
\DeclareMathOperator{\diag}{diag}
\newcommand{\Eins}{\ensuremath{\mathbbm{1}}}

\begin{document}

%=====================START FEYNMF===========================
\newcommand{\fmfklecks}[1]
             {\fmfv{d.filled=30,d.shape=circle,d.size=3mm}{#1}}
\newcommand{\fmfkreuz}[1]
             {\fmfv{d.filled=0,d.shape=cross,d.size=3mm}{#1}}
\newcommand{\fmfstrom}[1]
             {\fmfv{d.shape=circle,d.filled=empty,d.size=3mm}{#1}}
\newcommand{\fmfx}[1]{\fmfipair{kreuz} \fmfiequ{kreuz}{vloc(__#1)}
        \fmffreeze \fmfcmd{vdraw; vinit;} 
        \fmfcmd{cfilldraw ( (polycross 4)  scaled 3mm shifted kreuz);}}
\begin{fmffile}{graphen}
% A-Boson: modifizierte dashes
\fmfcmd{style_def bosonA expr p =
 save dpp; numeric dpp;
 dpp = (ceiling (pixlen (p, 10) / dash_len)) / length p;
 cdraw point 0 of p .. point (.25)/dpp of p;
 for k=1 upto dpp*length(p)-1:
    cdraw point (k-.25)/dpp of p .. point (k+.25)/dpp of p;
 endfor
 cdraw point (dpp*length(p)-.25)/dpp of p .. point length(p) of p;
enddef;}
% B-Boson: wiggly
\fmfcmd{style_def bosonB expr p = draw_wiggly p; enddef;}
% Majorana-Fermion: plain
\fmfcmd{style_def majorana expr p = cdraw p; enddef;}

%%%%%%%%%%%%%% TITEL %%%%%%%%%%%%%%%%%%%%%%%%%%%%%%%%
\title{\bf\vspace{1cm} Finite-Temperature Supersymmetry:\\
           The Wess-Zumino Model}
\author {{Kai Kratzert} \\[3mm]
        {\normalsize\it Deutsches Elektronen-Synchrotron DESY}\\
        {\it\normalsize D--22603 Hamburg, Germany}}
\date{\normalsize March 31, 2003\\[2cm]}
\begin{titlepage}
\maketitle
\thispagestyle{empty}

\begin{abstract}
  We investigate the breakdown of supersymmetry at finite temperature.
  While it has been proven that temperature always breaks
  supersymmetry, the nature of this breaking is less clear. On the one
  hand, a study of the Ward-Takahashi identities suggests a
  spontaneous breakdown of supersymmetry without the existence of a
  Goldstino, while on the other hand it has been shown that in any
  supersymmetric plasma there should exist a massless fermionic
  collective excitation, the phonino.  Aim of this work is to unify
  these two approaches. For the Wess-Zumino model, it is shown that
  the phonino exists and contributes to the supersymmetric
  Ward-Takahashi identities in the right way displaying that
  supersymmetry is broken spontaneously with the phonino as the
  Goldstone fermion.
\end{abstract}
\vspace{-18.8cm}
\begin{flushright}
DESY 03--043 \\
hep-th/0303260\\
\end{flushright}
\vfill
\noindent
{\small e-mail:} {\small\tt kratzert@mail.desy.de} 
\end{titlepage}
%%%%%%%%%%%%%% TITEL SPEZIELL ELSEVIER %%%%%%%%%%%%%%%%%%%%%%%%%%%%%%%%

\section{Introduction}
\label{sec:intro}

Supersymmetry has become a central element in the extensions of the
standard model of elementary particle physics that currently attract
most attention.  It is not only an intriguing mathematical concept
with a distinct importance for theoretical physics as the only
nontrivial extension of the Poincar\'e algebra in relativistic quantum
field theory, but it has some very attractive properties also from the
point of view of phenomenology.

On the other hand, our insight into the evolution of the universe has
made an enormous progress over the last decades, and there has been a
stimulating interchange of ideas between cosmology and quantum field
theory. Of particular importance is the behaviour of symmetries in a
hot plasma like it is believed to have existed shortly after the big
bang, since phase transitions may have left their traces in the
present state of the universe. For example, it is well-known that the
spontaneously broken gauge symmetry of the standard model is restored
at high temperature, a phenomenon that appears quite generally for
global and gauge symmetries. So, if we believe that supersymmetry was
realized in the hot early stages of the universe in one or the other
way, it is indispensable to understand its behaviour at high
temperatures.

Considering the fact that this is such a fundamental problem, it is
astonishing how little has been done to solve it. Although it was
realized that supersymmetry generally breaks down at finite
temperature already in the earliest works on the
subject~\cite{Das:rx,Girardello:1980vv}, the subsequent literature
contains many contradictory statements, and our understanding of the
mechanism of supersymmetry breaking at finite temperature is still
unsatisfactory.

The current status can be summarized as follows. Only recently, it has
been rigorously proven that supersymmetry is always broken at any
finite temperature~\cite{Buchholz:1997mf}. In fact, this is not
surprising, as in thermal field theory the ground state is described
by a statistical ensemble with different populations of bosons and
fermions.  Since the thermal ground state is responsible for the
breakdown of supersymmetry, it has much in common with a spontaneous
breaking.  Thus, a natural question to ask is whether it is associated
with the existence of a massless Goldstone fermion.  The investigation
of the supersymmetric Ward-Takahashi
identities~\cite{Boyanovsky:1983tu,Matsumoto:im} showed that there
must be a zero-energy Goldstone mode.  However, since the rest frame
of the heat bath also breaks Lorentz invariance, this mode is not
necessarily associated with a propagating Goldstone particle.
Nevertheless, in~\cite{Lebedev:rz} (clarifying earlier ideas
in~\cite{Gudmundsdottir:1986uq}) it has been shown from a
complementary point of view that any model with thermally broken
supersymmetry should contain a massless fermionic collective
excitation, similar to the appearance of sound waves in a medium with
spontaneously broken Lorentz invariance. The associated particle was
baptized phonino because of its similarity to the phonon.  For the
simple case of the Wess-Zumino model, the existence of this phonino
was proved.

In this work, we will again focus on the Wess-Zumino model. Though it
appears relatively simple and has already been investigated in a
number of papers, it reveals a very interesting structure. Our aim is
to unify the different results in the literature and to obtain a
complete picture of the breakdown of supersymmetry in this model. To
this end, we will give an explicit proof of the existence of the
phonino and investigate its contributions to the Ward-Takahashi
identities of broken supersymmetry.  It turns out that supersymmetry
is indeed broken spontaneously by the heat bath with the phonino
playing the role of the Goldstone particle.  The results obtained for
the Wess-Zumino model allow us to infer the behaviour of more general
models.

The paper is organized as follows. After briefly reviewing the
theoretical framework and the Wess-Zumino model in
sections~\ref{sec:tft} and~\ref{sec:model}, the established knowledge
about supersymmetry and one-loop behaviour of the Wess-Zumino model
will be presented and partly extended for our purposes in
sections~\ref{sec:susy} and~\ref{sec:loop}.  In
section~\ref{sec:phonino}, we present a full calculation of the
fermion propagator for low momenta and give an explicit proof of the
existence of the phonino.  Finally, section~\ref{sec:wti} is devoted
to the investigation of the Ward-Takahashi identities.

%=============THERMAL FIELD THEORY=================================
\section{Thermal field theory}
\label{sec:tft}

We will work in the framework of finite temperature field theory which
is the appropriate formalism for the description of quantum fields in
thermal equilibrium. The thermal background will be described by the
canonical ensemble with a density matrix
\[  \rho=Z^{-1}\e^{-\beta H}. \]
Here, $H$ is the Hamiltonian, and the inverse temperature
\mbox{$\beta=(k_\mathrm{B} T)^{-1}$} is chosen in units so that
Boltzmann's constant is unity.  The partition function $Z$ normalizes
the density operator so that the expectation value of an
observable~$\mathcal{O}$ is given by
\[   \langle \mathcal{O}\rangle_\beta= \tr \rho\mathcal{O}. \]

The basic effect of the presence of the thermal background on free
quantum fields is a modification of the propagators. Since
annihilation operators do not annihilate the thermal ground state, the
propagator of a scalar field $A$ gets an additional thermal
contribution. It reads
\begin{equation}
\label{eq:propagator}
 D(p)=\intvierx \e^{ip(x-y)}\langle T A(x)A(y)\rangle_\beta=
 \frac{i}{p^2-m^2+i\epsilon}+2\pi\,\delta(p^2-m^2)\,\nb{p_0},
\end{equation}
where
\begin{equation}
  \nb{p_0}=\frac{1}{\e^{\beta |p_0|}-1}
\end{equation}
is the Bose-Einstein distribution function.  Analogously, the free
fermion propagator is given by
\begin{equation}
  S(p)=\intvierx \e^{ip(x-y)}\langle T \Psi(x)\Psibar(y)\rangle_\beta=
  \frac{i(\pslash+m)}{p^2-m^2+i\epsilon}
  -(\pslash+m)\,2\pi\,\delta(p^2-m^2)\,\nf{p_0},
\end{equation}
with the Fermi-Dirac distribution function
\begin{equation}
  \nf{p_0}=\frac{1}{\e^{\beta |p_0|}+1}.
\end{equation}
Already at this point it becomes clear that supersymmetry has a hard
time at finite temperature since the different distribution functions
lead to quite different thermal contributions to the bosonic and
fermionic propagators.

The treatment of an interacting theory requires much more effort. From
the two formalisms that have been developed for this purpose, we will
make use of the real-time formalism since it is suited for the direct
perturbative calculation of thermal correlation functions in real time
by standard Feynman diagram techniques.  For consistency, this
formalism requires a doubling of the degrees of freedom. Formally, one
introduces for each field a ghost field with the same interaction as
the original field, only of opposite sign. The thermal propagator of a
scalar field is then given by a matrix of the form
\begin{equation}
\label{eq:matrix_prop}
  \widetilde{\!D}(p)=M(p_0)\left(\begin{matrix}
     \dfrac{i}{p^2-m^2+i\epsilon}&\hphantom{-}0\\
     0&-\dfrac{i}{p^2-m^2-i\epsilon}\end{matrix}\right)M(p_0),
\end{equation}
where
\[
M(p_0)=\left(\begin{matrix}\cosh\theta(p_0)&\sinh\theta(p_0)\\
\sinh\theta(p_0)&\cosh\theta(p_0)\end{matrix}\right)
\quad \mathrm{with}\quad \sinh^2\theta(p_0)=\nb{p_0}.
\]
The 11-component of this matrix gives back the thermal
propagator~\eqref{eq:propagator}. More general correlation functions
can be calculated perturbatively by evaluating the same diagrams as in
the vacuum theory, where external lines always correspond to physical
1-fields.  It will later turn out that the ghost fields do not lead to
relevant contributions in our calculations, but neverless must be
considered in a consistent treatment.

The matrix structure~\eqref{eq:matrix_prop} allows also the
understanding of self energy corrections. Without going into any
details (for which we refer to the standard literature
as~\cite{Landsman:uw}), we note that the full propagator must have a
similar structure, and the thermal self energy corrections only lead
to the usual shift
\[ p^2-m^2\to p^2-m^2-\Pi_\beta(p), \]
while the overall structure~\eqref{eq:propagator} of the thermal
propagator remains unchanged.  The real part of the thermal self
energy function $\Pi_\beta(p)$ is directly accessible to a
perturbative calculation.  It coincides with the real part of the self
energy function with physical external lines, while the imaginary part
requires minor corrections.

In case the self energy is real and small compared to the mass, it
only induces a small, temperature-dependent mass shift.  One should
however note that there is now an ambiguity in defining the mass.
Because of the breakdown of Lorentz invariance in the heat bath, the
self energy $\Pi_\beta(p)$ not only depends on $p^2$ but also on
$\pvec$.  Therefore, the value of the propagator at zero momentum or
its pole for vanishing three-momentum lead to different notions of
mass while they coincide at zero temperature. For reasons that will
become clear later, we will adopt the definition of mass as the pole
of the full propagator in the limit of vanishing three-momentum,
\begin{equation}
  m^2_\beta=m^2+\Pi_\beta(m,0).
\end{equation}

The fermionic case appears quite similar. Here, the thermal
self energy leads to a temperature-dependent shift
\[  \pslash-m\to \pslash-m-\Sigma_\beta(p)  \]
so that the thermal mass is given by
\begin{equation}
m^2_\beta=\frac{m}{2}\tr\bigl[(1+\gamma^0)\Sigma_\beta(m,0)\bigr],
\end{equation}
as long as the self energy is small.  In general, the propagator can
of course have a much more complicated structure and in particular
involve an imaginary damping part.

In finite temperature field theory, ultraviolet divergences can be
absorbed by the same redefinition of the parameters as in the vacuum
theory so that higher order vacuum contributions are small after
renormalization.  In this work, we will only be concerned with the
additional thermal contributions.

%=======================WZ MODEL==================================
\section{The Wess-Zumino model}
\label{sec:model}

The Wess-Zumino model~\cite{Wess:tw} is the simplest supersymmetric
quantum field theory. It describes a single self-interacting chiral
superfield
\[ 
\Phi=\phi+\sqrt{2}\,\theta\psi+\theta\theta F
\]
whose component fields are a scalar field $\phi$, a Weyl fermion
$\psi$ and an auxiliary scalar field $F$.  The Lagrangian reads
\[ 
\mathcal{L}=\Phi^\dagger\Phi
|_{\theta\theta\shortoverline{\theta}\shortoverline{\theta}}
+\left(W(\Phi)|_{\theta\theta\vphantom{\shortoverline{\theta}}}
  +\mathrm{h.c.}\right)
\]
where the interaction is determined by the superpotential which we
will take as
\begin{equation}
\label{eq:superpot}
W(\Phi)= \frac{m}{2}\Phi^2+\frac{g}{3}\Phi^3.
\end{equation}

Without dwelling on the superspace formalism, we rewrite the
Lagrangian in terms of the component fields.  After eliminating the
auxiliary field through its algebraic equation of motion, the on-shell
Lagrangian reads
\begin{equation}
\label{eq:lagrangian}
\begin{split}
  \mathcal{L}&=\tfrac{1}{2}\Psibar (i \dslash-m) \Psi
  +\tfrac{1}{2}(\partial_\mu A\, \partial^\mu\! A-m^2 A^2)
  +\tfrac{1}{2}(\partial_\mu B\, \partial^\mu B-m^2 B^2)\\
  &-gmA(A^2+B^2)-\tfrac{1}{2}g^2(A^2+B^2)^2 
  -g\Psibar(A-i\gamma^5 B)\Psi.
\end{split}
\end{equation}
This Lagrangian describes a scalar field $A$ and a pseudoscalar field
$B$, defined as the real components of the scalar field $\phi$,
\begin{equation}
\label{eq:skalar}
  A=\tfrac{1}{\sqrt{2}}\,(\phi+\phi^\dagger),\quad
  B=-\tfrac{i}{\sqrt{2}}(\phi-\phi^\dagger),
\end{equation}
in interaction with the Majorana fermion
\begin{equation}
\label{eq:majorana}
 \Psi=\binom{\psi}{\shortoverline{\psi}}.
\end{equation}
As a consequence of supersymmetry, all fields have equal mass $m$, and
all couplings are determined by the parameters $g$ and $m$.  We assume
the coupling $g$ to be small, in order to allow a perturbative
treatment.

The propagation of the three fields in a thermal background can be
described by the usual real-time propagators for massive particles,
\begin{equation}
\begin{split}
D_{A/B}(q)&=\dfrac{i}{q^2-m^2+i\epsilon}+2\pi\delta(q^2-m^2)\nb{q_0},\\
S(q)&=\left(\qslash+m\right)\left(\dfrac{i}{q^2-m^2+i\epsilon}
-2\pi\delta(q^2-m^2)\nf{q_0}\right),
\end{split}
\end{equation}
where we will denote, in a diagrammatic language, $A$-bosons by dashed
lines and $B$-bosons by wiggly lines, while solid lines stand for the
fermion.

%=======================SUSY=========================================
\section{Supersymmetry}
\label{sec:susy}

The action formed by the Lagrangian~\eqref{eq:lagrangian} is invariant
under the on-shell supersymmetry transformations
\begin{equation}
\begin{split}
\label{eq:susy_wz}
  \delta A&=\shortoverline{\xi} \Psi \\
  \delta B&=\shortoverline{\xi} i\gamma^5 \Psi\\
  \delta \Psi&=-(i\dslash+m)(A+i\gamma^5 B)\xi
                -g(A+i\gamma^5B)^2\xi, 
\end{split}
\end{equation}
where $\xi$ is an infinitesimal fermionic transformation parameter.
These continuous transformations are associated with a conserved
current $J^\mu$, the supercurrent. It is given by~\cite{Sohnius:qm}
\begin{equation}
\label{eq:supercurrent}
  J^\mu=-(\dslash+im)A\gamma^\mu\Psi 
        +(\dslash-im)i\gamma^5B\gamma^\mu\Psi
        -ig(A+i\gamma^5B)^2\gamma^\mu\Psi.
\end{equation}
The supercharge
\[  Q=\intdreix J^0(x),  \]
lets us express the above supersymmetry transformations as the
commutator (for bosonic operators) or anticommutator (for fermionic
ones) with the charge,
\[  \delta\mathcal{O}=-i\shortoverline{\xi} [Q,\mathcal{O}]_\pm. \]

In terms of the supercharge, the supersymmetry algebra reads
\begin{equation}
\label{eq:susy_charge}
  \{Q,\shortoverline{Q}\}=2\gamma^\mu P_\mu,
\end{equation}
where $P_\mu$ is the energy-momentum operator. This relation with
Poincar\'e symmetry displays why supersymmetry is necessarily broken
in any thermal background. The nonvanishing energy density of the heat
bath, \mbox{$\langle P_0\rangle_\beta\neq 0$}, breaks Lorentz
invariance spontaneously and, by relation~\eqref{eq:susy_charge}, also
supersymmetry breaks down.

\subsection{Ward-Takahashi identities and Goldstone's theorem}
\label{sec:goldstone}

The existence of a conserved current is a highly nontrivial fact which
leads to important relations between time-ordered correlation
functions involving the symmetry current.  The general form of these
Ward-Takahashi identities reads
\begin{equation}
\label{eq:wt_allg}
 \partial^x_\mu\langle T J^\mu(x) 
 \mathcal{O}_1(y_1)\dots \mathcal{O}_n(y_n)\rangle 
 =\sum_{i=1}^n\deltav(x-y_i)\langle T \mathcal{O}_1(y_1)\dots
 \left[Q,\mathcal{O}_i(y_i)\right]_\pm \dots\mathcal{O}_n(y_n)\rangle.
\end{equation}
It is important to note that these identities are basically operator
identities, so that they are valid in the vacuum as well as in a
thermal state, as it was established in~\cite{Boyanovsky:1983tu}.
Furthermore, they are valid even if supersymmetry is broken
spontaneously. In the case of an explicit breaking, there is obviously
no reason to expect their validity. Thus, the Ward-Takahashi
identities provide a useful tool for the investigation of broken
symmetries.

The spontaneous breakdown of supersymmetry is characterized by some
ferm\-ionic operator $\mathcal{O}$ transforming inhomogeneously,
\[ \langle\left\{Q,\mathcal{O}(y)\right\}\rangle\neq 0. \]
The corresponding Ward-Takahashi identity,
\[
  \partial^x_\mu\langle T J^\mu(x)\mathcal{O}(y)\rangle 
  =\deltav(x-y)\langle\left\{Q,\mathcal{O}(y)\right\}\rangle,
\]
can be rewritten in momentum space. By defining
\[
  \Gamma^\mu_{J\mathcal{O}}(k)=\intvierx \e^{i k (x-y)}
\langle T J^\mu(x)\mathcal{O}(y)\rangle,
\]
one obtains
\[
  -i k_\mu\Gamma^\mu_{J\mathcal{O}}(k)
  =\langle\left\{Q,\mathcal{O}(y)\right\}\rangle\neq 0.
\]
In order to satisfy this equation for all momenta $k$, the Fourier
transformed correlation function~$\Gamma^\mu_{J\mathcal{O}}$ on the
left hand side necessarily has a pole for \mbox{$k=0$}. This is of
course nothing but Goldstone's theorem.  In the vacuum, it follows
from Lorentz invariance that there must be a pole for all lightlike
momenta \mbox{$k^2=0$}, that is, there must exist a massless Goldstone
particle.  One cannot however draw this conclusion at finite
temperature. In this case, Lorentz invariance is broken and there can
in principle be an isolated pole for vanishing momentum without the
need for a Goldstone particle.  This observation lead the authors
of~\cite{Boyanovsky:1983tu,Matsumoto:im} to the conclusion that the
thermal breakdown of supersymmetry is not associated with the
existence of a Goldstone particle. A similar observation has been made
for the related breakdown of Lorentz invariance~\cite{Ojima:1985is}.
In any case, the identification of a propagating Goldstone particle in
a specific model requires some more effort.

For our case of the Wess-Zumino model, the simplest Ward-Takahashi
identity one can consider is that for a single fermion operator,
\begin{equation}
\label{eq:wt_1}
  \partial^x_\mu\langle T J^\mu(x)\Psibar(y) \rangle_\beta =
  -i\deltav(x-y)m\veva.
\end{equation}
On tree level, this identity is clearly fulfilled in a trivial way.
The behaviour in the interacting theory is less trivial and will be
discussed in section~\ref{sec:wti}.

Secondly, we can establish the identity for the composite mode
$A\Psi$.  Inserting the supersymmetry
transformations~\eqref{eq:susy_wz} into the general
formula~\eqref{eq:wt_allg}, we obtain
\begin{equation}
\begin{split}
\label{eq:wt_2}
\partial^x_\mu\langle T J^\mu(x)A(y)\Psibar(z)\rangle_\beta
&=\deltav(x-y)\,i\langle T \Psi(y)\Psibar(z)\rangle_\beta\\
&\quad +\deltav(x-z)
(\dslash_y-im)\langle T A(y)A(z)\rangle_\beta+\mathcal{O}(g).
\end{split}
\end{equation}
It is worth studying how this identity is satisfied at finite
temperature.  Let us switch off the interaction for a while and
consider the special case \mbox{$y=z$}. In momentum space, the right
hand side then reads
\begin{eqnarray}
\label{eq:rhs}
\intvierx \e^{ik(x-y)}  RHS &=&
\intvier{q}\bigl(iS(q)+(-i\qslash-im)D_A(q)\bigr)\nonumber\\
 &=&-2\pi i\,m\intvier{q} \bigl(\nf{q_0}+\nb{q_0}\bigr)\delta(q^2-m^2).
\end{eqnarray}
As expected, the vacuum contributions from bosonic and fermionic
propagators cancel because of the equality of their masses. Thus, the
right hand side is trivial at zero temperature. At finite temperature,
in contrast, the different thermal propagators for boson and fermion
no longer cancel but leave a nontrivial right hand side. In the limits
of nonrelativistic and relativistic temperatures, it is
straightforward to calculate,
\begin{equation}
\label{eq:wt2_rhs}
\intvierx \e^{ik(x-y)}  RHS = \left\{\ \begin{array}{ll}
 -2i\e^{-m/T}\left(\dfrac{Tm}{2\pi}\right)^{3/2} &\quad T\ll m,\\
 -im\dfrac{T^2}{8}&\quad T\gg m.  \end{array} \right. 
\end{equation}
Thereby we have neglected any corrections suppressed by factors of
order \mbox{$T/m$} and \mbox{$m/T$}, respectively, which will always
be done in the following, as long as they do not become relevant.

The left hand side of~\eqref{eq:wt_2} gives
\begin{align}
\label{eq:lhs_apsi}
  LHS&=\partial^x_\mu \langle T (-\dslash_x-im)A(x)
       \gamma^\mu\Psi(x)A(y)\Psibar(y)\rangle_\beta\\
     &=\intvier{k} \e^{-ik(x-y)}\!\intvier{q}
       (i\kslash-i\qslash-im)D_A(k-q)(-i\kslash)S(q).\nonumber
\end{align}
The loop integral gets two contributions from the thermal part of
either of the propagators. The knowledge that these have support on
the mass shell simplifies the algebra considerably, and one calculates
\begin{align*}
  \intvierx \e^{ik(x-y)}  LHS&= \intvier{q} 
\frac{i(\kslash-\qslash-m)\kslash(\qslash+m)}{(k-q)^2-m^2+i\epsilon}
(-2\pi)\nf{q_0}\delta(q^2-m^2)\\ &\quad+\intvier{q} 
\frac{i(\qslash-m)\kslash(\kslash-\qslash+m)}{(k-q)^2-m^2+i\epsilon}
2\pi\nb{q_0}\delta(q^2-m^2)\\
&=-2\pi i\,m\intvier{q} \bigl(\nf{q_0}+\nb{q_0}\bigr)\delta(q^2-m^2).
\end{align*}
So, we are left with the same result as for the right hand
side~\eqref{eq:rhs}, and we have shown that the Ward-Takahashi
identity~\eqref{eq:wt_2} is fulfilled in a nontrivial way.  In the
vacuum, this would be possible only if there was a massless Goldstone
particle generating the pole in the correlation function on the left
hand side.  At finite temperature, the situation is different.
Although there is obviously no Goldstone fermion in the free, massive
theory, nevertheless the identity is non-trivially satisfied.

The identity for the mode $B\Psi$ behaves in the same way. One has
\begin{equation}
  \partial^x_\mu\langle T J^\mu(x)B(y)\Psibar(y)\rangle_\beta
  =\partial^x_\mu\langle T 
     J^\mu(x)A(y)\Psibar(y)\rangle_\beta\, i\gamma^5.
\end{equation}
Comparing the relative couplings of the two modes to the supercurrent,
one can identify the linear combination
\begin{equation}
\chi=A\Psi-i\gamma^5B\Psi 
\end{equation}
as the Goldstone mode. This is not surprising once rewritten in terms
of the original chiral superfield by means of
equations~\eqref{eq:skalar} and~\eqref{eq:majorana},
\begin{equation}
(A-i\gamma^5B)\Psi=\binom{\Phi^2|_{\theta\vphantom{\shortoverline{\theta}}}}{\smash{\Phi^\dagger}^2|_{\shortoverline{\theta}}}.
\end{equation}
Thus, the Goldstone mode is the fermionic component field of the
composite superfield $\Phi^2$ that determines the mass term in the
superpotential~\eqref{eq:superpot}. Note that the
result~\eqref{eq:wt2_rhs} is nontrivial only because of this mass
term.

It will turn out in section~\ref{sec:wti} that in the full interacting
theory the Ward-Takahashi identities are satisfied in a somewhat
different way. It will be shown that the Goldstone mode indeed
corresponds to a propagating particle. Furthermore, the Goldstone mode
$\chi$ interacts with the fundamental fermion by means of the Yukawa
interaction
\[  \mathcal{L}_Y=-g\Psibar(A-i\gamma^5 B)\Psi. \]
Thereby, a Goldstone pole will appear also in the fermion propagator
which reflects itself in a nontrivial behaviour of the Ward-Takahashi
identity~\eqref{eq:wt_1}.

So far, our discussion overlaps with earlier results
from~\cite{Boyanovsky:1983tu}. Let us go one step further and see what
we can learn for the general case. Clearly, the way the first
identity~\eqref{eq:wt_1} is satisfied is highly dependent on the
details of the model. The second identity~\eqref{eq:wt_2} is more
universal since it involves, to leading order, only the mass and
behaves non-trivially already at the tree level.  However, we would
rather expect a nontrivial behaviour completely independent of masses
and couplings since, for the reasons given above, the thermal
breakdown of supersymmetry is a universal phenomenon.  According to
the supersymmetry algebra~\eqref{eq:susy_charge}, it is tied to that
of Lorentz invariance.  Consequently, the most general Ward-Takahashi
identity one should consider is that for the supercurrent itself. Its
transformation law is given by
\begin{equation}
 \label{eq:susy_current}
 \{Q, \Jbar^{\mu}\}=2\,T^{\mu\nu}\gamma_\nu+\dots,
\end{equation}
where the additional terms vanish when taken as the expectation value
in a translation invariant state like the vacuum or a thermal
equilibrium state~\cite{Sohnius:qm}.  This is basically the local form
of~\eqref{eq:susy_charge} as the energy-momentum tensor $T^{\mu\nu}$
is the current generating the translations $P^\mu$.  The corresponding
Ward-Takahashi identity then reads
\begin{equation}
\label{eq:wt_3}
  \partial^x_\mu\langle T J^\mu(x)\Jbar^\nu(y) \rangle
  =\deltav(x-y)\, 2\langle T^{\nu\mu}\rangle\gamma_\mu
\end{equation}
and is completely independent of the model.  The nonvanishing of the
right hand side is characteristic to any thermal background, and the
identity implies that a nontrivial Goldstone mode must appear in all
operators contributing to the supercurrent. In this way, the
identity~\eqref{eq:wt_2} is rather a special case since, in the
massive theory, the bilinear operator $A\Psi$ is part of the
supercurrent~\eqref{eq:supercurrent}.

Let us verify the identity~\eqref{eq:wt_3} for our case of the
Wess-Zumino model.  The right hand side involves the thermal
expectation value of the energy-momentum tensor. In a thermal
equilibrium state, it reads
\[ \langle T^{\mu\nu}\rangle_\beta=\diag(\rho,p,p,p), \]
where the energy density $\rho$ is given by
\[ 
 \rho=\intdrei{q} E_q \bigl(2\nf{E_q}+2\nb{E_q}\bigr)= 
  \left\{\ \begin{array}{ll} 
  4m\e^{-m/T}\left(\dfrac{Tm}{2\pi}\right)^{3/2}&\quad T\ll m,\\[2mm]
    \dfrac{\pi^2}{8}T^4 &\quad T\gg m. \end{array}\right.
\]
The pressure $p$ is related with the energy density as
\[
p=\left\{\ \begin{array}{ll} 
\dfrac Tm\, \rho&\quad T\ll m,\\[3mm]
\dfrac 13\, \rho &\quad T\gg m. \end{array}\right.
\]
Hence, on the right hand side of the identity~\eqref{eq:wt_3} we
find, for $T\ll m$,
\begin{equation}
\label{eq:t_low}
 2\langle T^{\nu\mu}\rangle_\beta\,\gamma_\mu=8m\e^{-m/T}
\left(\frac{Tm}{2\pi}\right)^{3/2}\left(\begin{array}{c}\gamma_0\\[1.5mm]
 \tfrac{T}{m}\,\vec\gamma\end{array} \right).
\end{equation}
For $T\gg m$, we have
\begin{equation}
\label{eq:t_high}
2\langle T^{\nu\mu}\rangle_\beta\,\gamma_\mu
=\frac{\pi^2}{4}T^4\left(\begin{array}{c}\gamma_0\\[1.5mm]
      \tfrac{1}{3}\,\vec\gamma\end{array} \right).
\end{equation}

This must be compared with the left hand side of~\eqref{eq:wt_3},
\begin{equation*}
\begin{split}
  LHS&=\partial^x_\mu \langle T (-\dslash_x-im)A(x)\gamma^\mu\Psi(x)
       \Psibar(y)\gamma^\nu(-\dslash_y+im)A(y)\rangle_\beta\\ 
       &\quad +\partial^x_\mu \langle T (\dslash_x-im)i\gamma^5B(x)
       \gamma^\mu\Psi(x)\Psibar(y)\gamma^\nu 
       i\gamma^5(\dslash_y+im)B(y)\rangle_\beta.\\
\end{split}
\end{equation*}
It can be evaluated in the same manner as
equation~\eqref{eq:lhs_apsi}.  The same cancellations take place, and
one obtains, after some algebra,
\[
 \intvierx \e^{ik(x-y)}  LHS
 =4\intvier{q} q^\nu \qslash\, 2\pi
 \bigl(\nf{q_0}+\nb{q_0}\bigr)\delta(q^2-m^2).
\]
Calculating the remaining integral, one ends up with exactly the same
result as for the right hand side, equations~\eqref{eq:t_low}
and~\eqref{eq:t_high}.  Thus, the Ward-Takahashi
identity~\eqref{eq:wt_3} is shown to be fulfilled in a nontrivial way.

\subsection{Supersymmetric sound}
\label{sec:supersound}

In view of the fact that the quantum field theoretical framework not
necessarily requires a propagating Goldstone particle associated with
the thermal breakdown of supersymmetry, it is even more interesting
that the existence of a massless fermionic excitation can be deduced
in a complementary setting.  In~\cite{Lebedev:rz} (see
also~\cite{Leigh:1995jw}), a hydrodynamic approach to the
supersymmetric plasma has lead to the prediction of a slow-moving
collective excitation whose existence should be as general as that of
sound waves.

\begin{sloppypar} 
Consider a relativistic perfect fluid with an energy-momentum tensor
of the form
\[ \langle T^{\mu\nu}\rangle_\beta=\diag(\rho,p,p,p). \]
  If the system is disturbed by a small, spacetime-dependent variation
  of, say, the temperature, $\Delta T(x)$, the conservation of the
  energy-momentum tensor, \mbox{$\partial_\mu\langle\delta
    T^{\mu\nu}\rangle_\beta=0$}, translates to a wave equation for the
  disturbance,
\begin{equation}
\left(\frac{\partial \rho}{\partial T}\,\partial_0^2-
\frac{\partial p}{\partial T}\,
\smash{\vec{\mspace{.5mu}\partial}}^2\right)\Delta T(x)=0.
\end{equation}
It describes the propagation of sound waves with the velocity
\begin{equation}
v^2_S=\frac{\partial p/ \partial T}{\partial \rho/ \partial T}.
\end{equation}
Thus, as a consequence of the breakdown of Lorentz symmetry in the
thermal bath, small perturbations in the energy-momentum tensor
propagate as sound waves. Their quanta, the phonons, can be viewed as
the Goldstone bosons associated with the spontaneous symmetry
breaking.
\end{sloppypar}

Now, in the supersymmetric case, one can imagine the system undergoing
a small, spacetime-dependent supersymmetry variation $\xi(x)$.  The
conservation of the supercurrent, \mbox{$\partial_\mu \langle \delta
  J^\mu(x)\rangle_\beta=0$}, translates to a wave equation for the
transformation parameter,
\begin{equation}
(\rho\,\gamma_0\partial_0+p\,\vec\gamma\vec{\mspace{.5mu}\partial})
\,\xi(x)=0,
\end{equation}
where the components of the energy-momentum tensor enter through the
transformation law~\eqref{eq:susy_current}.  This Dirac equation
describes a massless fermionic excitation propagating with the
velocity
\begin{equation}
v_{SS}=\frac{p}{\rho}.
\end{equation}
Thus, in a system whose supersymmetry is broken by the thermal bath,
there should exist `supersymmetric sound waves'. Their quanta,
interpreted as the Goldstone fermions associated with the spontaneous
symmetry breaking, are naturally called phoninos.

Both sound and supersymmetric sound have a very characteristic
dispersion law given by
\begin{equation}
v_S^2=v_{SS}=\left\{\ \begin{array}{ll} \frac Tm&\quad T\ll m\\[1.5mm]
      \frac 13 &\quad T\gg m \end{array}\right.
\end{equation}
in the non-relativistic and relativistic limits.

Despite the similarity to sound waves, there is surely no classical
picture of these fermionic waves. It should however be possible to
verify their existence in the framework of thermal field theory.  Just
as sound waves appear as poles in the correlator~\mbox{$\langle
  T_{\mu\nu}(x)T_{\rho\sigma}(y)\rangle_\beta$}, supersymmetric sound
must lead to poles in the correlation function~\mbox{$\langle
  J_{\mu}(x)\Jbar_{\!\nu}(y)\rangle_\beta$}.  Since supersymmetric
sound appears as a collective phenomenon, it is expected to appear in
the quantum field theoretical framework only as a nonperturbative
effect.

%------------------Abschnitt: Ein-Loop---------------------------------
\section{One-loop corrections}
\label{sec:loop}

Though the thermal breakdown of supersymmetry is evident already at
the tree level, a deeper understanding certainly requires the
investigation of higher-order effects. The one-loop effects have
already been investigated in earlier works, so that we can refer
to~\cite{Midorikawa:1984fi,Lebedev:rz} for explicit calculations.

Firstly, the interaction with the heat bath leads to the appearance of
a nontrivial expectation value of the fundamental scalar field.  Its
value is given by
\begin{equation}
\label{eq:vev}
\veva=\left\{\ \begin{array}{ll} -gm\alpha&\quad T\ll m,\\[1mm] 
      -\dfrac{gT^2}{2m} &\quad T\gg m, \end{array}\right.
\end{equation}
where we have introduced the small parameter
\begin{equation}
\label{eq:alpha}
  \alpha=8m^{-3}\intdrei{q} \e^{-(m+\frac{\qvec^2}{2m})/T}
 =\e^{-m/T}\left(\frac{2T}{\pi m}\right)^{3/2}.
\end{equation}
The nonzero value is another clear sign for the breakdown of
supersymmetry. It must be stressed, however, that it is not a
necessary condition like in the vacuum. In the massless Wess-Zumino
model, for example, the expectation value vanishes because of chiral
symmetry while supersymmetry is nevertheless broken for the general
reasons given above.

Secondly, the interaction leads to a modification of the propagators.
The dominant self energy corrections come from the proper one-loop
diagrams and, with a contribution of the same order of magnitude, from
the coupling to the thermal expectation value of the scalar field.
For momenta \mbox{$p=(m,0)$}, the relevant real parts of the thermal
self energy functions $\Pi^\beta_i(p)$ have been calculated
in~\cite{Lebedev:rz}.  Their result can easily be generalized to
momenta on the mass shell, since none of the diagrams shows a
dependence on the three-momentum~$\pvec$, as long as \mbox{$p^2=m^2$}.
As a consequence, the self energy corrections only lead to a small
displacement of the masses,
\[  m_i^2=m^2+\bigl.\Pi^\beta_i(p)\bigr|_{p^2=m^2}  \]
without changing the dispersion relation
\[  p_0^2=\pvec^2+m_i^2. \]
Explicitly, the values of the effective thermal masses are given by
\begin{equation}
\label{eq:massen}
T\ll m:\, \left\{\,
\begin{array}{rl}
  m_A^2&=m^2-\tfrac{20}{3}g^2\alpha m^2\\[1mm]
  m_B^2&=m^2\\[1mm]
  m_\Psi^2&=m^2-2 g^2\alpha m^2
\end{array}\right.
\qquad
T\gg m:\, \left\{\,
\begin{array}{rl}
  m_A^2&=m^2-2g^2 T^2\\[1mm]
  m_B^2&=m^2\\[1mm]
  m_\Psi^2&=m^2-g^2 T^2
\end{array}\right.
\end{equation}
As yet another evidence for the breakdown of supersymmetry, we find
that the mass degeneracy of the three fields is lifted by the one-loop
thermal self energy corrections. As expected, the mass splitting
vanishes in the limit of zero temperature when supersymmetry is
restored.

In a model where supersymmetry is broken spontaneously already in the
vacuum, the masses of the members of one supermultiplet are related by
the mass formula
\[   \sum_J (-1)^{2J} (2J+1)\,m_J^2=0. \]
For the effective thermal masses~\eqref{eq:massen}, we find that this
relation is no longer valid, at least for low temperatures where
\mbox{$m_A^2+m_B^2\neq 2m_\Psi^2$}.  This raises the question whether
the breakdown of supersymmetry should at all be considered
spontaneous or rather explicit.

%------------------Abschnitt:Phonino-----------------------------------
\section{The phonino}
\label{sec:phonino}

\begin{figure}[t]
\begin{center}
\begin{fmfgraph*}(70,60)
  \fmfforce{0w,.2h}{i1} \fmfforce{1w,.2h}{o1}
  \fmfforce{.5w,.2h}{v1} \fmfforce{.5w,.7h}{v2} \fmfkreuz{v2}
  \fmf{majorana}{i1,v1,o1} \fmf{bosonA}{v1,v2}
\end{fmfgraph*}
\qquad
\begin{fmfgraph*}(100,60)
  \fmfleft{i1} \fmfright{o1} \fmf{majorana,tension=2.7}{i1,v1}
  \fmf{majorana,left}{v2,v1} \fmf{majorana,tension=2.7}{v2,o1}
  \fmf{bosonA,right}{v2,v1}
\end{fmfgraph*}
\qquad
\begin{fmfgraph*}(100,60)
  \fmfleft{i1} \fmfright{o1} \fmf{majorana,tension=2.7}{i1,v1}
  \fmf{majorana,left}{v2,v1} \fmf{majorana,tension=2.7}{v2,o1}
  \fmf{bosonB,right}{v2,v1}
\end{fmfgraph*}
\caption{One-loop contributions to the fermion self energy}
\label{fig:oneloop_f}
\end{center}
\end{figure}
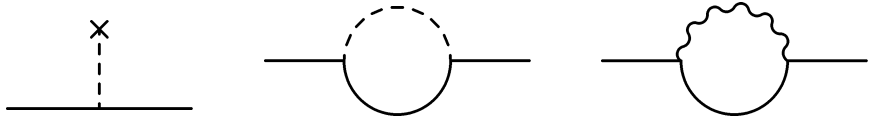

In the previous section, we have seen that the influence of the heat
bath on the fundamental fields is basically a small shift in their
masses.  According to the general arguments given in
section~\ref{sec:susy}, however, we suspect the existence of
additional fermionic excitations.  Hence, we should take a closer look
at the full self energy function that determines the poles of the full
fermion propagator. These are characterized by the condition
\begin{equation}
\label{eq:pole_cond}
\det\mathcal{S}^{-1}(p)=-i\det\bigl(\pslash-m-\Sigma_\beta(p)\bigr)=0.
\end{equation}
To leading order, the self energy $\Sigma_\beta$ gets contributions
from the coupling to the expectation value of the scalar field $A$ as
well as from the proper one-loop diagrams as drawn in
figure~\ref{fig:oneloop_f}.  Altogether, one finds
\begin{equation}
\label{eq:olse}
 -i\,\Sigma_\beta(p)=
  -2ig\veva-4g^2\intvier{q} D(p-q)
  \left(S(q)+i\gamma^5S(q)i\gamma^5\right).
\end{equation}
We are particularly interested in the limit of small
momenta.\pagebreak[3] Thus, we set the three-momentum to zero and
calculate
\[
\Sigma_\beta(p_0,0)=2g\veva+\frac{8g^2\gamma^0}{p_0}
\intdrei{q}\left(\frac{\nb{E_q}}{E_q}\frac{2E_q^2-p_0^2}{4E_q^2-p_0^2}
  +\frac{\nf{E_q}}{E_q}\frac{2E_q^2}{4E_q^2-p_0^2}\right),
\]
where $E_q=\sqrt{\smash[t]{\qvec}^2+m^2}$.
One observes that the second part develops a pole for vanishing
momentum.  The reason is of course the degeneracy of bosonic and
fermionic masses. In the limit \mbox{$p\to 0$}, the poles of both
propagators in the loop coincide, leading to an anomalously big
self energy.

In the relevant limits of low and high temperatures, the integration
can easily be performed. For small momenta $p_0$, the result can be
expressed in terms of~$\veva$ as
\[  \Sigma_\beta(p_0,0)=2g\veva-\frac{mg\veva}{p_0}\gamma^0. \]
Inserting this result into condition~\eqref{eq:pole_cond} for the
propagator poles, one obtains the dispersive equation
\[
 \left(p_0+\frac{mg\veva}{p_0}\right)^2=\bigl(m+2g\veva\bigr)^2
\]
with the solutions
\begin{align*}
  |p_0|&=m+g\veva+\mathcal{O}(g^2\veva^2),\\
  |p_0|&=g\veva+\mathcal{O}(g^2\veva^2).
\end{align*}
The first solution reproduces the small shift in the fermion mass that
was already calculated in the previous section. The existence of a
second solution indicates another excitation with a tiny mass.  In
fact, such a pole is what we expect, albeit with exactly vanishing
mass.  It thus seems that our simple one-loop calculation is not
sufficient.

As it will turn out in the following sections, a consistent
calculation of the full propagator requires to take into account the
full propagators in the internal lines, that is, to consider the
differences in the effective masses of bosons and fermion due to the
breakdown of supersymmetry. In this way, one must perform a
nonperturbative, selfconsistent calculation in order to uncover the
desired massless pole in the fermion propagator.

\subsection{The case of high temperature}

Let us now analyze the fermion self energy more carefully. At first,
we restrict ourselves to the relativistic case.  By the assumption
that the mass shifts are small compared to the mass, our discussion is
then limited to a temperature range \mbox{$m\ll T\ll g^{-1}m$}.

Since the interaction with the heat bath modifies the propagators by a
small, constant mass correction, we will approximate the
full propagators by the one-loop resummed propagators
\begin{equation}
\label{eq:fullprops}
\begin{split}
  \mathcal{S}(q)&=\left(\qslash+m+\Sigma'(q)\right)
  \left(\frac{i}{q^2-m_\Psi^2+i\epsilon}
    -2\pi\delta(q^2-m_\Psi^2)\nf{q_0}\right)\\
  \mathcal{D}_{A/B}(q)&=\frac{i}{q^2-m_{A/B}^2+i\epsilon}
  +2\pi\delta(q^2-m_{A/B}^2)\nb{q_0}
\end{split}
\end{equation} 
with the effective thermal masses as given in
equation~\eqref{eq:massen}.  As one can check from
equation~\eqref{eq:olse}, the one-loop self energy $\Sigma_\beta$
splits into a constant part~$\Sigma_\Eins$ proportional to the unit
matrix and a traceless part $\Sigma_\gamma$ (which is an odd function
of the momentum), so that one has
\[  \Sigma'(q)=\Sigma_\Eins-\Sigma_\gamma(q). \]

With this approximation for the full propagators, we can now evaluate
the fermion self energy~\eqref{eq:olse}.  For small momenta $k$, the
product of both propagators in the loop can be approximated as
\begin{equation}
\label{eq:produkt}
  \mathcal{D}_i(k-q)\mathcal{S}(q)=
  -2\pi i\left(\qslash+m+\Sigma'(q)\right)
  \delta(q^2-m^2)\frac{\nb{q_0}+\nf{q_0}}{m_\Psi^2-m_i^2-2qk},
\end{equation}
where we only keep corrections linear in $k$ to small terms of order
$g^2$. A general loop integral then reads, for
sufficiently regular $f$,
\begin{multline}
\label{eq:loopint}
  \intvier{q} \mathcal{D}_i(k-q)\mathcal{S}(q)f(q)\\
=-i\intdrei{q}\sum_{q_0=\pm E_q}\frac{\nb{E_q}+\nf{E_q}}{2E_q}
\frac{\qslash+m+\Sigma'(q)}{m_\Psi^2-m_i^2-2qk}
f(q).
\end{multline}
One finds that the small mass splitting between bosons and fermion
regularizes the loop integral. So, in contrast to the one-loop
calculation with tree-level masses, the integral no longer diverges in
the limit~\mbox{$k=0$} but becomes anomalously large because of the
small mass difference appearing in the denominator.  We assume that
\mbox{$qk\ll m_\Psi^2-m_i^2$} with a typical loop momentum
\mbox{$q\sim T$} which restricts ourselves to momenta \mbox{$k\ll
  g^2T$}.

The explicit calculation of the full self energy~\eqref{eq:olse} then
gives
\begin{equation*}
\begin{split}
  \Sigma_\beta(k)&=-\frac{4g^2}{m_\Psi^2-m_A^2}(m+\Sigma_\Eins)
  \intdrei{q}\frac{\nb{E_q}+\nf{E_q}}{E_q}\\
  &\quad-\frac{4g^2}{m_\Psi^2-m_B^2}(-m-\Sigma_\Eins)
  \intdrei{q}\frac{\nb{E_q}+\nf{E_q}}{E_q}\\
  &\quad-8g^2\left(k_0\gamma_0+\tfrac{1}{3}\vec k\vec\gamma\right)
  \left(\frac{1}{(m_\Psi^2-m_A^2)^2}
        +\frac{1}{(m_\Psi^2-m_B^2)^2}\right)
  \cdot\\&\quad\quad\cdot
  \intdrei{q}E_q\bigl(\nb{E_q}+\nf{E_q}\bigr)\\
  &\quad+\Sigma_\Eins.
\end{split}
\end{equation*}
One can now insert the mass splittings resulting from
equation~\eqref{eq:massen},
\[
  {m_\Psi^2-m_A^2}={m_B^2-m_\Psi^2}={g^2T^2},
\]
and perform the integration. One ends up with the simple result
\[
\Sigma_\beta(k)=-m-\frac{\pi^2}{g^2}
          \left(k_0\gamma_0+\tfrac{1}{3}\vec k\vec\gamma\right).
\]
Remarkably, the result is exactly the negative Lagrangian mass so
that, in our approximation, the full inverse fermion propagator reads
\begin{equation}
\label{eq:prop_gg}
  i\mathcal{S}^{-1}(k)=\kslash-m-\Sigma_\beta(k)\approx
\frac{\pi^2}{g^2}\left(k_0\gamma_0+\tfrac{1}{3}\vec k\vec\gamma\right).
\end{equation}
Because of the cancellation of the mass term, we find that the full
fermion propagator has an additional pole for vanishing momentum as
well as on the dispersion curve
\begin{equation}
  k_0=\pm\frac{1}{3}|\vec k|.
\end{equation}
This displays the existence of a massless excitation propagating with
velocity \mbox{$v=\frac{1}{3}$}, just as one expects from the general
arguments given in section~\ref{sec:supersound}.  Consequently, we
identify this pole in the fermion propagator as the phonino.

It should be added that this mechanism to cancel the fermion mass by
loop corrections goes back to a widely unnoticed paper by
Kapusta~\cite{Kapusta:1982eg} who performed a similar calculation for
the somewhat simpler case of the massless Wess-Zumino model.  However,
the author attributed the existence of the soft fermion to chiral
symmetry and not to the breakdown of supersymmetry.

\subsection{The case of low temperature}
\label{sec:low}

The above calculation can in principle be translated also to the case
of low temperature. It turns out, however, that this case requires a
more thorough investigation. We have seen that contributions of the
form~\eqref{eq:produkt} from two propagators of nearly degenerate mass
and opposite momentum lead to an anomalously large value of the self
energy. So, for consistency, one must take care of all diagrams of a
similar structure that could give equally large contributions.

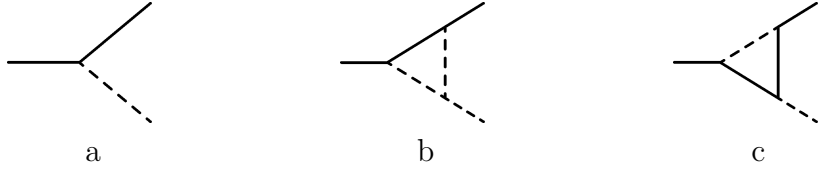
\begin{figure}[t]
\begin{center}
\begin{tabular}{ccc}
    \fmfframe(25,10)(25,0){\begin{fmfgraph}(60,45)
      \fmfleft{i1} \fmfright{o1,o2}  
      \fmf{majorana,tension=2}{i1,v1} \fmf{majorana}{v1,o2}
      \fmf{bosonA}{v1,o1}
    \end{fmfgraph}}
&
    \fmfframe(25,10)(25,0){\begin{fmfgraph}(60,45)
      \fmfleft{i1} \fmfright{o1,o2} 
      \fmf{majorana,tension=2.5}{i1,v1}\fmf{majorana}{v1,v2}
      \fmf{majorana,tension=1.5}{v2,o2} \fmf{bosonA}{v1,v3} 
      \fmf{bosonA,tension=1.5}{v3,o1}
      \fmf{bosonA,tension=0}{v3,v2}
    \end{fmfgraph}}
&
    \fmfframe(25,10)(25,0){\begin{fmfgraph}(60,45)
      \fmfleft{i1} \fmfright{o1,o2} 
      \fmf{majorana,tension=2.5}{i1,v1}\fmf{majorana}{v1,v2}
      \fmf{majorana,tension=0}{v2,v3}\fmf{majorana,tension=1.5}{v3,o2} 
      \fmf{bosonA}{v1,v3} \fmf{bosonA,tension=1.5}{v2,o1}
    \end{fmfgraph}}
\\
a&b&c
\end{tabular}
\end{center}
\caption{Tree-level Yukawa coupling and one-loop corrections}
\label{fig:vertex}
\end{figure}

Take a look at the vertex corrections in figure~\ref{fig:vertex}.  For
vanishing external momentum and low temperature, the contribution from
diagram~b is of order
\[
  g^3m\intdrei{q}\frac{\nf{E_q}+\nb{E_q}}{E_q}\frac{m}{m_\Psi^2-m_A^2}
 \frac{1}{m^2} \sim g.
\]
Hence, it is as relevant as the tree-level coupling.
For high temperature, in contrast, one finds
\[
  g^3m\intdrei{q}\frac{\nf{E_q}+\nb{E_q}}{E_q}\frac{T}{m_\Psi^2-m_A^2}
  \frac{1}{T^2} \sim \frac{m}{T}g,
\]
so it does not give a relevant contribution because of the tree-level
Yukawa coupling being proportional to \mbox{$m\ll T$}.  The same is
true for diagrams with fermion exchange like in
figure~\ref{fig:vertex}\,c.  Thus, for high temperature there is no
need to take these diagrams into account, while for low temperatures
one must evaluate all relevant contributions to the vertex. 

Obviously, many-loop diagrams play an equally important role, as long
as each intermediate vertex or line is again followed by a pair of
particles with a tiny mass difference, which leads to a cancellation
of the small coupling constants and distribution functions that
otherwise suppress any higher-order diagram.  Altogether, one must sum
up all ladder diagrams of this kind.

The standard way to perform such a resummation is by solving the
corresponding Bethe-Salpeter equations. Their general form reads,
diagrammatically,
\[
    \parbox{16mm}{\begin{fmfgraph}(40,35)
        \fmfleft{i1} \fmfright{o2,o1}
        \fmfklecks{v1}
        \fmf{majorana,tension=1.6}{i1,v1} \fmf{majorana}{v1,o1}
        \fmf{bosonA}{o2,v1}
    \end{fmfgraph}}
    =\  \parbox{16mm}{\begin{fmfgraph}(40,35)
        \fmfleft{i1} \fmfright{o2,o1}
        \fmf{majorana,tension=1.6}{i1,v1} \fmf{majorana}{v1,o1}
        \fmf{bosonA}{o2,v1}
    \end{fmfgraph}}
    +\
    \parbox{24mm}{\begin{fmfgraph}(70,35)
        \fmfleft{i1} \fmfright{o1,o2}
        \fmfklecks{v1}
        \fmfpolyn{empty,tension=0.7}{m}{4} \fmf{majorana}{i1,v1}
        \fmf{majorana,left=.7,tension=0.4}{v1,m4}
        \fmf{majorana,tension=.8}{o2,m3}
        \fmf{bosonA,right=.7,tension=0.4}{v1,m1}
        \fmf{bosonA,tension=.8}{m2,o1}
    \end{fmfgraph}}.
\]
Here, the shaded spot stands for the full proper (one-particle
irreducible) vertex while the square denotes any proper (two-particle
irreducible) scattering subdiagram, and a sum over all possible
intermediate states is implied. The internal lines stand for full
propagators.

In our specific case, the intermediate state in the Bethe-Salpeter
equation is always a boson-fermion pair of nearly degenerate mass. The
two-particle irreducible scattering amplitude gets contributions from
both boson and fermion exchange. So, the Bethe-Salpeter equation for
the vertex function involving the $A$~boson reads
\begin{equation}
\begin{split}
  \parbox{20mm}{\begin{fmfgraph}(50,40) \fmfleft{i1} \fmfright{o2,o1}
      \fmfklecks{v1}
      \fmf{majorana,tension=1.6}{i1,v1} \fmf{majorana}{v1,o1}
      \fmf{bosonA}{o2,v1}
    \end{fmfgraph}}
  =\ \parbox{20mm}{\begin{fmfgraph}(50,40) \fmfleft{i1}
      \fmfright{o2,o1} \fmf{majorana,tension=1.6}{i1,v1}
      \fmf{majorana}{v1,o1} \fmf{bosonA}{o2,v1}
    \end{fmfgraph}}
  &+\ \parbox{23mm}{\begin{fmfgraph}(65,40) \fmfleft{i1}
      \fmfright{o1,o2}
      \fmfklecks{v1}
      \fmf{majorana,tension=1.3}{i1,v1}
      \fmf{majorana,tension=0.5}{v2,v1} \fmf{majorana}{o2,v2}
      \fmf{bosonA,tension=0.5}{v3,v1} \fmf{bosonA}{o1,v3}
      \fmf{bosonA,tension=0}{v2,v3}
    \end{fmfgraph}}
  +\ \parbox{23mm}{\begin{fmfgraph}(65,40) \fmfleft{i1}
      \fmfright{o1,o2}
      \fmfklecks{v1}
      \fmf{majorana,tension=1.3}{i1,v1}
      \fmf{majorana,tension=0.5}{v1,v2} \fmf{majorana}{v3,o2}
      \fmf{bosonA,tension=0.5}{v3,v1} \fmf{bosonA}{o1,v2}
      \fmf{majorana,tension=0}{v2,v3}
    \end{fmfgraph}}\\[1mm]
  &+\ \parbox{23mm}{\begin{fmfgraph}(65,40) \fmfleft{i1}
      \fmfright{o1,o2}
      \fmfklecks{v1}
      \fmf{majorana,tension=1.3}{i1,v1}
      \fmf{majorana,tension=0.5}{v1,v2} \fmf{majorana}{o2,v2}
      \fmf{bosonB,tension=0.5}{v3,v1} \fmf{bosonA}{o1,v3}
      \fmf{bosonB,tension=0}{v2,v3}
    \end{fmfgraph}}
  +\ \parbox{23mm}{\begin{fmfgraph}(65,40) \fmfleft{i1}
      \fmfright{o1,o2}
      \fmfklecks{v1}
      \fmf{majorana,tension=1.3}{i1,v1}
      \fmf{majorana,tension=0.5}{v1,v2} \fmf{majorana}{v3,o2}
      \fmf{bosonB,tension=0.5}{v3,v1} \fmf{bosonA}{o1,v2}
      \fmf{majorana,tension=0}{v2,v3}
    \end{fmfgraph}}
\end{split}
\end{equation}    
A similar equation must hold for the $B$ boson vertex,
\begin{equation}
\begin{split}
  \parbox{20mm}{\begin{fmfgraph}(50,40) \fmfleft{i1} \fmfright{o2,o1}
      \fmfklecks{v1}
      \fmf{majorana,tension=1.6}{i1,v1} \fmf{majorana}{v1,o1}
      \fmf{bosonB}{o2,v1}
    \end{fmfgraph}}
  =\ \parbox{20mm}{\begin{fmfgraph}(50,40) \fmfleft{i1}
      \fmfright{o2,o1} \fmf{majorana,tension=1.6}{i1,v1}
      \fmf{majorana}{v1,o1} \fmf{bosonB}{o2,v1}
    \end{fmfgraph}}
  &+\ \parbox{23mm}{\begin{fmfgraph}(65,40) \fmfleft{i1}
      \fmfright{o1,o2}
      \fmfklecks{v1}
      \fmf{majorana,tension=1.3}{i1,v1}
      \fmf{majorana,tension=0.5}{v1,v2} \fmf{majorana}{o2,v2}
      \fmf{bosonA,tension=0.5}{v3,v1} \fmf{bosonB}{o1,v3}
      \fmf{bosonB,tension=0}{v2,v3}
    \end{fmfgraph}}
  +\ \parbox{23mm}{\begin{fmfgraph}(65,40) \fmfleft{i1}
      \fmfright{o1,o2}
      \fmfklecks{v1}
      \fmf{majorana,tension=1.3}{i1,v1}
      \fmf{majorana,tension=0.5}{v1,v2} \fmf{majorana}{v3,o2}
      \fmf{bosonA,tension=0.5}{v3,v1} \fmf{bosonB}{o1,v2}
      \fmf{majorana,tension=0}{v2,v3}
    \end{fmfgraph}}\\[1mm]
  &+\ \parbox{23mm}{\begin{fmfgraph}(65,40) \fmfleft{i1}
      \fmfright{o1,o2}
      \fmfklecks{v1}
      \fmf{majorana,tension=1.3}{i1,v1}
      \fmf{majorana,tension=0.5}{v1,v2} \fmf{majorana}{o2,v2}
      \fmf{bosonB,tension=0.5}{v3,v1} \fmf{bosonB}{o1,v3}
      \fmf{bosonA,tension=0}{v2,v3}
    \end{fmfgraph}}
  +\ \parbox{23mm}{\begin{fmfgraph}(65,40) \fmfleft{i1}
      \fmfright{o1,o2}
      \fmfklecks{v1}
      \fmf{majorana,tension=1.3}{i1,v1}
      \fmf{majorana,tension=0.5}{v1,v2} \fmf{majorana}{v3,o2}
      \fmf{bosonB,tension=0.5}{v3,v1} \fmf{bosonB}{o1,v2}
      \fmf{majorana,tension=0}{v2,v3}
    \end{fmfgraph}}
\end{split}
\end{equation}    

Let us translate these diagrammatic equations into mathematical
language.  
The amputated full proper vertex functions
will be denoted by~$\Gamma_{\!A}(k,p)$ and $\Gamma_{\!B}(k,p)$,
respectively,
\[ \parbox{3cm}{\begin{fmfgraph*}(50,40) 
    \fmfleft{i1}\fmflabel{$k$}{i1} \fmfright{o2,o1}
    \fmfv{label=$p$,label.angle=0}{o1}
    \fmfv{label=$k-p$,label.angle=0}{o2} \fmfklecks{v1}
    \fmf{majorana,tension=1.6}{i1,v1} \fmf{majorana}{v1,o1}
    \fmf{bosonA}{o2,v1}
\end{fmfgraph*}}
\equiv\Gamma_{\!A}(k,p),\quad\qquad
\parbox{3cm}{\begin{fmfgraph*}(50,40) \fmfleft{i1}\fmflabel{$k$}{i1}
    \fmfright{o2,o1} \fmfv{label=$p$,label.angle=0}{o1}
    \fmfv{label=$k-p$,label.angle=0}{o2} \fmfklecks{v1}
    \fmf{majorana,tension=1.6}{i1,v1} \fmf{majorana}{v1,o1}
    \fmf{bosonB}{o2,v1}
\end{fmfgraph*}}
\equiv\Gamma_{\!B}(k,p), \] which we want to calculate for small
momenta $k$.  The thermal contributions to the Bethe-Salpeter
equations come from loop momenta on the mass shell. Therefore, we
consider only on-shell momenta \mbox{$p^2=m^2$}.

All contributions to the Bethe-Salpeter equations have a very similar
structure.  Approximating the full propagators in the internal lines
by the one-loop resummed propagators~\eqref{eq:fullprops}, we again
find expressions of the form~\eqref{eq:loopint}.  For simplicity, we
introduce
\begin{equation}
\label{eq:def_delta}
  \Delta_A=m_\Psi^2-m_A^2=\frac{14}{3}g^2\alpha m^2,\qquad
  \Delta_B=m_\Psi^2-m_B^2=-2g^2\alpha m^2.
\end{equation}
Like in the case of high temperatures, we only consider momenta $k$
small compared to the mass differences and keep only the linear terms,
while small corrections of order $\Delta_i/m$ or $k/m$ will be
neglected.  Furthermore, we can safely neglect contributions of order
$\pvec/m$ or $\qvec/m$ (where $q$ is the momentum of the fermion in
the loop), since~$\pvec$ and $\qvec$ are thermal momenta of order
$\sqrt{Tm}$ which, for low temperature, is negligible compared to the
mass.  The propagators of the exchanged particles are then essentially
constant but depend on the relative sign of the zero components of
outgoing and loop momenta.

Altogether, the Bethe-Salpeter equation for the $A$ boson vertex can
be written as
\begin{equation}
\begin{split}
  \Gamma_{\!A}(k,p)&=-2ig-4g^2\intdrei{q}\frac{\e^{-E_q/T}}{E_q}\cdot\\
  &\quad\cdot\Bigg\{
  \left(\frac{\pslash+2m}{3m^2}+\frac{3m}{-m^2}\right)
  \left.\frac{\qslash+m}{\Delta_A-2qk}
    \Gamma_{\!A}(k,q)\right|_{q_0=E_q\cdot\sign{p_0}} \\
  &\qquad+\left. \left(\frac{\pslash+2m}{-m^2}+\frac{3m}{3m^2}\right)
    \left.\frac{\qslash+m}{\Delta_A-2qk}
      \Gamma_{\!A}(k,q)\right|_{q_0=-E_q\cdot\sign{p_0}} \right.\\
  &\qquad-\left. \left(\frac{-\pslash+2m}{3m^2}+\frac{m}{-m^2}\right)
    \left.\frac{-\qslash+m}{\Delta_B-2qk}
   i\gamma^5\Gamma_{\!B}(k,q)\right|_{q_0=E_q\cdot\sign{p_0}}\right.\\
  &\qquad- \left(\frac{-\pslash+2m}{-m^2}+\frac{m}{3m^2}\right)
  \left.\frac{-\qslash+m}{\Delta_B-2qk}
    i\gamma^5\Gamma_{\!B}(k,q)\right|_{q_0=-E_q\cdot\sign{p_0}}
  \Bigg\}.
\end{split}
\end{equation}
For the $B$ boson vertex, one finds
\begin{equation}
\begin{split}
  i\gamma^5\Gamma_{\!B}(k,p)&=
  -2ig-4g^2\intdrei{q}\frac{\e^{-E_q/T}}{E_q}\cdot\\
  &\quad\cdot\Bigg\{ \left(\frac{\pslash}{3m^2}+\frac{-m}{-m^2}\right)
  \left.\frac{\qslash+m}{\Delta_A-2qk}
    \Gamma_{\!A}(k,q)\right|_{q_0=E_q\cdot\sign{p_0}} \\
  &\qquad+\left. \left(\frac{\pslash}{-m^2}+\frac{-m}{3m^2}\right)
    \left.\frac{\qslash+m}{\Delta_A-2qk}
      \Gamma_{\!A}(k,q)\right|_{q_0=-E_q\cdot\sign{p_0}} \right.\\
  &\qquad-\left. \left(\frac{-\pslash}{3m^2}+\frac{m}{-m^2}\right)
    \left.\frac{-\qslash+m}{\Delta_B-2qk}
    i\gamma^5\Gamma_{\!B}(k,q)\right|_{q_0=E_q\cdot\sign{p_0}}\right.\\
  &\qquad-\left. \left(\frac{-\pslash}{-m^2}+\frac{m}{3m^2}\right)
    \frac{-\qslash+m}{\Delta_B-2qk}
    i\gamma^5\Gamma_{\!B}(k,q)\right|_{q_0=-E_q\cdot\sign{p_0}}
  \Bigg\}.
\end{split}
\end{equation}

These seemingly complex coupled integral equations can be simplified
considerably by introducing the functions
\begin{equation*}
  \tilde{\Gamma}_{\!A}(k,p)=
   \frac{\pslash+m}{\Delta_A-2pk}\,\Gamma_{\!A}(k,p),\quad
  \tilde{\Gamma}_{\!B}(k,p)=
    -i\gamma^5\frac{\pslash+m}{\Delta_B-2pk}\,\Gamma_{\!B}(k,p).
\end{equation*}
Rewriting the Bethe-Salpeter equations in terms of these functions, it
turns out that the right hand sides involve only the symmetric parts
\begin{equation}
\label{eq:def_vab}
  V_{A,B}(k,p) =\left(\tilde{\Gamma}_{A,B}(k,p)
      +\tilde{\Gamma}_{A,B}(k,-p)\right)_{p_0=E_p}.
\end{equation}
By straightforward calculation, one can express the Bethe-Salpeter
equations in terms of the functions $V_{A,B}$ as
\begin{equation}
\label{eq:bsgtilde}
\begin{split}
V_A(k,p)&
  =\left(\frac{2m}{\Delta_A}+\frac{4\pslash\,pk}{\Delta_A^2}\right)
\left\{-2ig+\frac{4g^2}{m^2}\intdrei{q} \e^{-E_q/T}
\left( 2 V_A(k,q) + \frac{2}{3} V_B(k,q)\right)\right\} \\
V_B(k,p)&
  =\left(-\frac{2m}{\Delta_B}+\frac{4\pslash\,pk}{\Delta_B^2}\right)
\left\{-2ig+\frac{4g^2}{m^2}\intdrei{q} \e^{-E_q/T}
\left(- \frac{2}{3} V_A(k,q) + \frac{2}{3} V_B(k,q) \right) \right\},
\end{split}
\end{equation}
where we have again made a linear approximation for small momenta $k$.

For vanishing $k$, one finds that the functions $V_{A,B}(0,p)$ do not
depend on the momentum $p$ at all. The integration can then easily be
performed by making use of the relations
\begin{equation*}
  \frac{4g^2}{m^2}\intdrei{q} \e^{-E_q/T}=\frac12g^2 \alpha m=
\frac{3}{28}\frac{\Delta_A}{m}=-\frac{1}{4}\frac{\Delta_B}{m}
\end{equation*}
which immediately follow from equations~\eqref{eq:alpha}
and~\eqref{eq:def_delta}. One arrives at the system of linear
equations
\begin{equation}
\label{eq:system}
\begin{split}
V_A(0,p)&=\frac{2m}{\Delta_A}\left\{-2ig+\frac{3}{28}\frac{\Delta_A}{m}
\left(2V_A(0,p)+\frac{2}{3}V_B(0,p)\right)\right\}\\
V_B(0,p)&=-\frac{2m}{\Delta_B}\left\{-2ig-\frac{1}{4}\frac{\Delta_B}{m}
\left(-\frac{2}{3}V_A(0,p)+\frac{2}{3}V_B(0,p)\right)\right\}
\end{split}
\end{equation}
which is easy to solve. The solution is given by 
\begin{equation}
\label{eq:vab}
V_A(0,p)=V_B(0,p)=\frac{2m}{\Delta_B}\cdot 2ig.
\end{equation}
It is illustrative to express this result in terms of the amputated
vertex functions  $\Gamma_{\!A}$ and $\Gamma_{\!B}$ we started with.
One finds
\begin{equation}
\Gamma_{\!B}(0,p)=-2ig\left(-i\gamma^5\right),\quad
\Gamma_{\!A}(0,p)=-2ig\left(-\frac{\Delta_A}{\Delta_B}\right).
\end{equation}
In other words, the Yukawa coupling to the $B$ boson comes out of the
whole resummation without a change while the coupling to the $A$ boson
is enhanced by a factor of
\[ -\frac{\Delta_A}{\Delta_B}=\frac{7}{3}. \]

Let us now calculate the momentum dependence of the vertex
functions. To this end, we expand in $k$,
\[ V_{A,B}(k,p)=V_{A,B}(0,p)+k_\mu C^{\mu}_{A,B}(p), \]
which is substituted in the Bethe-Salpeter
equations~\eqref{eq:bsgtilde}.  The resulting integral equations for
the functions $C^{\mu}_{A,B}(p)$ can be simplified by replacing the
known values of $V_{A,B}(0,p)$ and integrating, so that one obtains
\begin{equation}
\label{eq:system_cab}
\begin{split}
  k_\mu C^\mu_A(p)&= 
    -\frac{14}{3}ig\frac{4\pslash\,pk}{\Delta_A^2} 
    +\frac{8g^2}{m\Delta_A}\intdrei{q} \e^{-E_q/T}
    \left( 2\,k_\mu C^\mu_A(q)+\frac{2}{3}\,k_\mu C^\mu_B(q) \right),\\
  k_\mu C^\mu_B(p)&=
    -2ig\frac{4\pslash\,pk}{\Delta_B^2}
    -\frac{8g^2}{m\Delta_B}\intdrei{q} \e^{-E_q/T}
    \left(-\frac{2}{3}\,k_\mu C^\mu_A(q)+\frac{2}{3}\,k_\mu C^\mu_B(q)
  \right).
\end{split}
\end{equation}
At this point, it is no longer possible to neglect the momentum
dependence since in particular the vector components $\vec C_{A,B}(p)$
receive an important contribution rising quadratically with $|\pvec|$.
Yet, the system can be solved by multiplication by $\e^{-E_p/T}$,
followed by integration. In this way, one finds a linear system with
the solution
\begin{equation}
\label{eq:cab}
  \intdrei{q} \e^{-E_q/T} k_\mu C^\mu_A(q) =\intdrei{q}
  \e^{-E_q/T} k_\mu C^\mu_B(q)
  =\frac{im^2}{2g\Delta_B}\left(m\,k_0\gamma_0+T\,\vec
  k\vec\gamma\right).
\end{equation}
One can now proceed and calculate the functions $C^\mu_{A,B}(q)$ by
inserting this result into equations~\eqref{eq:system_cab}.  We skip
this last step since the result will not be needed in the following.

Instead, we proceed to the calculation of the full fermion propagator.
With our knowledge of the full one-particle irreducible vertex
functions, we can now evaluate the full proper self energy.  In terms
of full propagators and vertex functions, the main contribution
from~\eqref{eq:olse} translates to
\begin{equation*}
-i\,\Sigma_\beta(k)=-2ig\intvier{q}\mathcal{D}(k-q)\left(\Gamma_A(k,q)
\mathcal{S}(q)+\Gamma_B(k,q) \mathcal{S}(q) i\gamma^5\right),
\end{equation*}
which, by making use of equations~\eqref{eq:loopint}
and~\eqref{eq:def_vab}, is easily brought into the form
\begin{equation*}
  \Sigma_\beta(k)=-2ig\intdrei{q}
  \frac{\e^{-E_q/T}}{E_q}\bigl(V_{A}(k,q)+V_{B}(k,q)\bigr).
\end{equation*}
With our results for $V_{A,B}$ in~\eqref{eq:vab} and~\eqref{eq:cab},
one calculates
\begin{equation}
  \Sigma_\beta(k)=-m-\frac{1}{g^2\alpha}
\left(k_0\gamma_0+\frac{T}{m}\vec k\vec\gamma\right).
\end{equation}
Thus, the self energy \mbox{$\Sigma_\beta(0)=-m$} exactly cancels the
Lagrangian mass term in the full inverse propagator which thereby
becomes
\begin{equation}
\label{eq:prop_ll}
 i \mathcal{S}^{-1}(k)=\kslash-m-\Sigma_\beta(k)\approx
\frac{1}{g^2\alpha}
\left(k_0\gamma_0+\frac{T}{m}\vec k\vec\gamma\right).
\end{equation}
So we have finally found the additional massless pole in the full
fermion propagator. It becomes singular for vanishing momentum as well
as on the dispersion curve
\begin{equation}
  k_0=\pm\frac{T}{m}|\vec k|
\end{equation}
which is nothing but the dispersion law predicted in
section~\ref{sec:supersound}. Therefore, this pole can be identified
with the phonino.  Thus we have shown that the phonino pole in the
fermion propagator is present at any moderate temperature, only the
residue and dispersion law change with temperature. In the limit
\mbox{$T\to 0$}, the residue vanishes and the Goldstone mode
disappears, since supersymmetry is restored.

\begin{sloppypar}
Due to our approximation, we have proven the existence of the phonino
pole only for small momenta. We have assumed that
\mbox{$kq\ll\Delta_A$} with a typical thermal momentum
\mbox{$|\qvec|\sim\sqrt{Tm}$}, so that the derivation is restricted to
momenta \mbox{$|\vec k|\ll{\Delta_A}/{\sqrt{Tm}}$}. Interestingly,
this is not only a technical point but of physical relevance. In the
hydrodynamic picture, the existence of sound and supersymmetric sound
waves is restricted to long wavelengths much greater than the mean
free path so that there is always time to establish local
thermodynamic equilibrium. At higher frequencies, the waves are
strongly damped as it was shown in~\cite{Gudmundsdottir:1986uq}.
\end{sloppypar}

The fact that our calculation requires a resummation of both one-loop
corrections to the propagators and higher-order vertex corrections
shows that the existence of the phonino is really a nonperturbative
phenomenon. This is consistent with the interpretation as a collective
excitation according to the hydrodynamic explanation given in
section~\ref{sec:supersound}. Furthermore, it explains why earlier
calculations were not able to relate the Goldstone mode to a
propagating excitation by simpler one-loop
calculations~\cite{Boyanovsky:1983tu,Matsumoto:im}.

The existence of the low-temperature phonino pole and dispersion law
was already shown in~\cite{Lebedev:rz} by investigating the
homogeneous Bethe-Salpeter equations for the full (i.e., not
one-particle irreducible) vertex functions. Our explicit calculation
of the full fermion propagator is however a new result which will be
essential for the following discussion of the Ward-Takahashi
identities.

%------------------Abschnitt:WTI--------------------------------------
\section{Verification of the Ward-Takahashi identities}
\label{sec:wti}

The appearance of the massless phonino derived in the previous section
is apparently linked with the breakdown of supersymmetry.  In order to
obtain a complete picture and to prove that the phonino is indeed the
Goldstone fermion of spontaneously broken supersymmetry, let us now
take a closer look at the supersymmetric Ward-Takahashi identities and
verify that the phonino contributes in exactly the way expected for a
Goldstone particle.

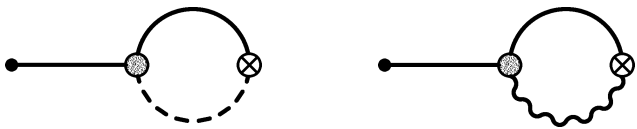
\begin{figure}[b]
\begin{center}
\begin{fmfgraph*}(90,60)
  \fmfstraight \fmfleft{o1} \fmfright{i1} \fmfdot{o1} \fmfklecks{v1}
  \fmfstrom{i1} \fmf{bosonA,left,width=.8thick}{i1,v1}
  \fmf{majorana,tension=1.8,width=.8thick}{v1,o1}
  \fmf{majorana,left,width=.8thick}{v1,i1} \fmfx{i1}
\end{fmfgraph*}
\qquad\qquad
\begin{fmfgraph*}(90,60)
  \fmfstraight \fmfleft{o1} \fmfright{i1} \fmfdot{o1} \fmfklecks{v1}
  \fmfstrom{i1} \fmf{bosonB,left,width=.8thick}{i1,v1}
  \fmf{majorana,tension=1.8,width=.8thick}{v1,o1}
  \fmf{majorana,left,width=.8thick}{v1,i1} \fmfx{i1}
\end{fmfgraph*}
\end{center}
\caption{Main contributions to the Ward-Takahashi 
identity~\eqref{eq:wteins}}
\label{fig:wt1}
\end{figure}

In section~\ref{sec:susy}, we have derived the identity
\begin{equation}
\label{eq:wteins}
\partial^x_\mu\langle T J^\mu(x)\Psibar(y) \rangle_\beta =
  -im\veva\,\deltav(x-y).
\end{equation}
At the one-loop level, the right hand side is nontrivial because of
the nonvanishing thermal expectation value of the scalar field given
in equation~\eqref{eq:vev}. Thus, the left hand side must get a
contribution from a Goldstone mode.  Surely, we expect the phonino to
act this part, with the main contributions coming from the class of
diagrams drawn in figure~\ref{fig:wt1}.  Here, the encircled cross
denotes the supercurrent~\eqref{eq:supercurrent}. The shaded spot
stands for the full vertex, and the thick lines mean full resummed
thermal propagators.

The evaluation of the these diagrams thus gives, to leading order,
\begin{multline}
\label{eq:wtleft}
  \langle T J^\mu(x)\Psibar(y)\rangle_\beta=\\
\begin{aligned}
  \quad&\intvier{k}
  \e^{-ik(x-y)}\Biggl\{\intvier{q}\left(i\kslash-i\qslash-im\right)
  \gamma^\mu\mathcal{D}_A(k-q)
  \mathcal{S}(q)\Gamma_{\!A}(k,q)\mathcal{S}(k)\\
  &\quad+\intvier{q}\left(-i\kslash+i\qslash-im\right)
        i\gamma^5\gamma^\mu
  \mathcal{D}_B(k-q)\mathcal{S}(q)\Gamma_{\!B}(k,q)
  \mathcal{S}(k)\Biggr\}.
\end{aligned}
\end{multline}

Going over to momentum space, we amputate the fermion propagator and
define
\[
  \Gamma_{J\smash[t]{\Psibar}}^\mu(k)=
  \intvierx \e^{ik(x-y)}\langle T J^\mu(x)\Psibar(y)\rangle_\beta\, 
  \mathcal{S}^{-1}(k)
\]
so that the Ward-Takahashi identity~\eqref{eq:wteins} we want to verify
can be written as
\begin{equation}
\label{eq:wt1_ft}
 -ik_\mu\Gamma_{J\smash[t]{\Psibar}}^\mu(k)=-im\veva\, 
 \mathcal{S}^{-1}(k).
\end{equation}

Now to the calculation of the left hand side. The loop integrals
in~\eqref{eq:wtleft} are again of the form~\eqref{eq:loopint}
well-known by now, and the Dirac algebra is simplified by the fact
that the loop-momentum is on-shell.  Since we are mainly interested in
the role of the Goldstone mode, we consider small momenta $k$ and only
keep the linear terms.  We find
\begin{equation*}
\begin{split}
-ik_\mu\Gamma_{J\smash[t]{\Psibar}}^\mu(k)
=-i\intdrei{q}\!\sum_{q_0=\pm E_q}\!&\Bigg\{\left(-2\qslash\,qk\right)
\frac{\nb{E_q}+\nf{E_q}}{2E_q}
\frac{\Gamma_{\!A}(0,q)}{m_\Psi^2-m_A^2}\\
&+2\qslash\,qk\frac{\nb{E_q}+\nf{E_q}}{2E_q}
\frac{i\gamma^5\Gamma_{\!B}(0,q)}{m_\Psi^2-m_B^2}\Bigg\}.
\end{split}
\end{equation*}
The integration gives, for nonrelativistic temperatures 
\mbox{$T\ll m$},
\begin{equation}
\label{eq:gamma_low}
-ik_\mu\Gamma_{J\smash[t]{\Psibar}}^\mu(k)=\frac{m^2}{g}
\left(k_0\gamma_0+\frac{T}{m}\vec k\vec\gamma\right).
\end{equation}
Similarly, for the relativistic case \mbox{$T\gg m$}, one finds
\begin{equation}
\label{eq:gamma_high}
-ik_\mu\Gamma_{J\smash[t]{\Psibar}}^\mu(k)
=\frac{\pi^2T^2}{2g}
\left(k_0\gamma_0+\frac{1}{3}\vec k\vec\gamma\right).
\end{equation}
One immediately observes the similarity to the inverse phonino
propagators obtained in equations~\eqref{eq:prop_gg}
and~\eqref{eq:prop_ll}, and indeed, one finds by comparison with
equation~\eqref{eq:vev} that the difference is just a factor of
\mbox{$-im\veva$}. So, the Ward-Takahashi identity~\eqref{eq:wt1_ft}
is fulfilled through the contribution from the phonino, and we can
conclude that the phonino is the Goldstone fermion associated with the
spontaneous breakdown of supersymmetry in the thermal background.

Once we have convinced ourselves that the Ward-Takahashi identity is
satisfied at finite temperature in the full resummed interacting
theory, we can in turn use it to obtain more information about the
full propagator. If one keeps also the linear terms in the calculation
of $\Gamma^\mu_{J\smash[t]{\Psibar}}(k)$,
equation~\eqref{eq:wtleft}, one can deduce the
nonlinear corrections to the phonino dispersion law. In agreement
with~\cite{Lebedev:rz}, they are found to be small in the region where
the phonino exists.

Let us now proceed to the second Ward-Takahashi identity, the one for
the composite mode $A\Psi$ that was derived in
equation~\eqref{eq:wt_2}. Its right hand side was evaluated already in
equation~\eqref{eq:wt2_rhs} so that we can rewrite the identity in
terms of $\veva$ as
\begin{equation}
\label{eq:wtzwei}
\partial^x_\mu\langle T J^\mu(x)A(y)\Psibar(y)\rangle_\beta 
=i\frac{m^2}{4g}\veva\,\deltav(x-y).
\end{equation}
Again, the right hand side is nontrivial at nonvanishing temperature,
so that the left hand side must get a contribution from the Goldstone
mode. 

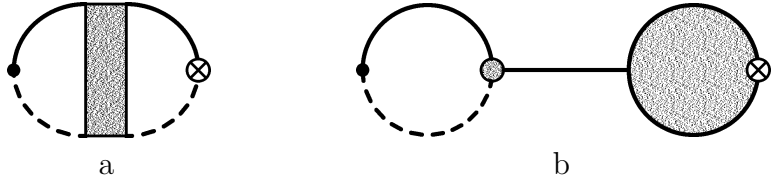
\begin{figure}[b]
\begin{center}
\begin{tabular}{cc}
\fmfframe(25,0)(25,0){\begin{fmfgraph*}(70,50)
 \fmfleft{i1} \fmfright{o1} \fmfstrom{o1}  \fmfpolyn{filled=30}{v}{4} 
 \fmfforce{.39w,0h}{v1} \fmfforce{.61w,0h}{v2} \fmfforce{.61w,1h}{v3} 
 \fmfforce{.39w,1h}{v4} \fmfposition
 \fmfi{majorana,width=.8thick}{vloc(__i1) {up} .. {right} vloc(__v4)}
 \fmfi{majorana,width=.8thick}{vloc(__v3) {right} .. {down} vloc(__o1)}
 \fmfi{bosonA,width=.8thick}{vloc(__i1) {down} .. {right} vloc(__v1)}
 \fmfi{bosonA,width=.8thick}{vloc(__v2) {right} .. {up} vloc(__o1)}
 \fmfdot{i1} \fmfx{o1}
\end{fmfgraph*}}
&
\fmfframe(25,0)(25,0){\begin{fmfgraph*}(150,50)
  \fmfstraight \fmfleft{o1} \fmfright{i1} \fmfdot{o1} \fmfklecks{v2}
  \fmfstrom{i1} \fmf{majorana,left,tag=1,width=.8thick}{i1,v1}
  \fmf{majorana,left,tag=2,width=.8thick}{v1,i1}
  \fmf{majorana,tension=1.9,width=.8thick}{v1,v2}
  \fmf{majorana,left,width=.8thick}{o1,v2}
  \fmf{bosonA,left,width=.8thick}{v2,o1} \fmfposition \fmfdraw
  \fmfipath{p[]} \fmfiset{p1}{vpath1(__i1,__v1) ..  vpath2(__v1,__i1)
  .. cycle} \fmfcmd{drawhalftone (.8 , p1 );} \fmfx{i1}
\end{fmfgraph*}}
\\
a&b
\end{tabular}
\end{center}
\caption{Contributions to the Ward-Takahashi 
identity~\eqref{eq:wtzwei}}
\label{fig:wt2}
\end{figure}

The diagrams contributing to the left hand side fall into two classes,
drawn in figure~\ref{fig:wt2}. For the first class of diagrams, one
must have control over the full one-particle irreducible four-point
function denoted by a shaded box that could in principle have
additional poles. However, in the limit of small momentum we are
interested in, it can easily be shown that this is not the case.  The
relevant contributions to this four-point function can be resummed by
means of the same Bethe-Salpeter equations that were set up for the
calculation of the one-particle irreducible three-point functions in
section~\ref{sec:low}. At this point, it is only important that this
resummation does not introduce any new poles, in other words that the
homogeneous Bethe-Salpeter equation does not have any nontrivial
solutions. This can easily be seen from equation~\eqref{eq:system}
which, as a homogeneous system, has only the trivial solution.  Hence,
this class of diagrams stays finite in the limit of low momentum.

Thus, the relevant contributions come from the second class of
diagrams involving the phonino, schematically drawn in
figure~\ref{fig:wt2}\,b.  Here, the big smudge stands for the sum of
all diagrams discussed above in connection with the first
Ward-Takahashi identity.  So, for the evaluation of this correlation
function in the low-momentum limit, we can take advantage of our
earlier results and write
\begin{equation}
\begin{split}
  \Gamma_{JA\smash[t]{\Psibar}}^\mu(k)&\equiv \intvierx
  \e^{ik(x-y)}\langle T J^\mu(x)A(y)\Psibar(y)\rangle_\beta\nonumber\\
  &=\Gamma_{J\smash[t]{\Psibar}}^\mu(k)\mathcal{S}(k) \intvier{q}
  \shortoverline{\Gamma}_{\!A}(k,q)\mathcal{D}_A(k-q)\mathcal{S}(q)
  \nonumber\\
  &=\Gamma_{J\smash[t]{\Psibar}}^\mu(k)\mathcal{S}(k)
  \left(-\frac{m}{4g}\right),
\end{split}
\end{equation}
where the numerical factor is the same for low and high temperatures.

We can now make use of the first Ward-Takahashi
identity~\eqref{eq:wt1_ft} and obtain
\begin{equation}
\label{eq:lhszwo}
-ik_\mu\Gamma_{JA\smash[t]{\Psibar}}^\mu(k)=i\frac{m^2}{4g}\veva
\end{equation}
which is nothing but the Ward-Takahashi identity~\eqref{eq:wtzwei} we
intended to prove, written in momentum space.

Let us compare the ways the Ward-Takahashi identity is fulfilled for
the free and interacting theories. In the free theory, the loop
diagram in figure~\ref{fig:wt2}\,a (without any interaction) is the
only one that contributes to the left hand side of the identity.
Because of the equality of the masses of both propagators, it becomes
singular in the vanishing momentum limit, thereby saturating the
Ward-Takahashi identity non-trivially without the need for a massless
mode.  In the interacting theory, in contrast, there appears a mass
splitting between fermion and bosons, so that diagram~\ref{fig:wt2}\,a
stays finite even in the limit of vanishing momentum. Hence, there
must be a massless Goldstone mode that contributes to the left hand
side in order to saturate the Ward-Takahashi identity. Our calculation
shows that the phonino pole in the fermion propagator exactly does
this job.

Finally, the Ward-Takahashi identity involving the supercurrent, 
\begin{equation}
\label{eq:wtdrei}
\partial^x_\mu\langle T J^\mu(x)\Jbar^\nu(y) \rangle_\beta
  =\deltav(x-y)\, 2\langle T^{\nu\mu}\rangle_\beta\,\gamma_\mu,
\end{equation}
 can be verified in the same way. The dominant contributions to the
 left hand side come again from diagrams with a single fermion
 intermediate state. Hence, we can write, in momentum space,
\begin{equation*}
\begin{split}
  \Gamma_{J\smash[t]{\Jbar}}^{\mu\nu}(k)&\equiv\intvierx 
  \e^{ik(x-y)}\langle T J^\mu(x)\Jbar^{\nu}(y)\rangle_\beta\\
  &=\Gamma_{J\smash[t]{\Psibar}}^\mu(k)\mathcal{S}(k)
  \shortoverline{\Gamma}_{J\smash[t]{\Psibar}}^\nu(k).
\end{split}
\end{equation*}
By making use of the Ward-Takahashi identity~\eqref{eq:wt1_ft}, we
find
\begin{equation*}
  -ik_\mu\Gamma_{J\smash[t]{\Jbar}}^{\mu\nu}(k)=-im\veva
  \shortoverline{\Gamma}_{J\smash[t]{\Psibar}}^\nu(k),
\end{equation*}
which can now be calculated by using our earlier results
on~$\Gamma^\mu_{J\smash[t]{\Psibar}}(k)$.

First, in the limit of low temperature, we make use
of~\eqref{eq:gamma_low} and obtain the left hand side of the
Ward-Takahashi identity as
\begin{equation*}
  -ik_\mu\Gamma_{J\smash[t]{\Jbar}}^{\mu\nu}(k)=
  -m^2\veva^2\,\frac{2m^2}{\Delta_B}\left(\begin{array}{c}
  \gamma_0\\[1mm] \tfrac{T}{m}\,\vec\gamma\end{array}\right)
  =8m\e^{-m/T}\left(\frac{Tm}{2\pi}\right)^{3/2}\left(
  \begin{array}{c}\gamma_0\\[1mm] 
  \tfrac{T}{m}\,\vec\gamma\end{array}\right).
\end{equation*}
For high temperatures, one calculates in the same way,
using~\eqref{eq:gamma_high},
\begin{equation*}
  -ik_\mu\Gamma_{J\smash[t]{\Jbar}}^{\mu\nu}(k)=m^2\veva^{\!2}\,
  \frac{\pi^2}{g^2}
  \left(\begin{array}{c}\gamma_0\\[1mm] \tfrac13\,\vec\gamma\end{array}
  \right)= \frac{\pi^2}{4}T^4\left(\begin{array}{c}\gamma_0\\[1mm]
  \tfrac13\,\vec\gamma\end{array} \right).
\end{equation*}
Comparing with our earlier results on the energy-momentum tensor,
equations~\eqref{eq:t_low} and~\eqref{eq:t_high}, we find that this
can be written as
\begin{equation}
  -ik_\mu\Gamma_{J\smash[t]{\Jbar}}^{\mu\nu}(k)= 2\langle
  T^{\nu\mu}\rangle_\beta\,\gamma_\mu.
\end{equation}
This is however nothing but the momentum space version of the
identity~\eqref{eq:wtdrei} we wanted to prove.

\section{Conclusion}
\label{sec:conc}

Our calculations have lead to a consistent picture of the behaviour of
the Wess-Zumino model at finite temperature. At moderate temperatures,
the interaction with the thermal background leads, on the one-loop
level, to a small splitting of the effective masses, and the scalar
field develops a nontrivial thermal expectation value. Both are signs,
if not necessary conditions, for the breakdown of supersymmetry.  As
an additional feature, the phonino pole appears in the propagator of
the fundamental fermion, as a full nonperturbative calculation shows.
The reason is the continuous interaction with boson-fermion pairs of
nearly degenerate mass. As a consequence, the same pole appears in all
modes that couple to the supercurrent. This relation to the
supercurrent indicates the role of the phonino as the Goldstone
particle of spontaneously broken supersymmetry, which could be
confirmed by the investigation of the Ward-Takahashi identities that
are saturated by the contribution from the phonino in a nontrivial
way.

However, our results are not restricted to the Wess-Zumino model.  The
general picture we have drawn and verified in this simple model,
according to which the breakdown of supersymmetry is a consequence of
the breakdown of Lorentz invariance, can immediately be translated to
any supersymmetric model.  Just as the existence of supersymmetric
sound was derived in~\cite{Lebedev:rz} in a model-independent way, the
Ward-Takahashi identity for the supercurrent generically predicts the
existence of a Goldstone pole in any mode that couples to the
supercurrent, as long as the fields involved contribute to the
nonvanishing energy density of the thermal bath.  By the same
mechanism that we explicitly studied for the Wess-Zumino model, this
pole must be associated with a propagating particle whenever the
interaction lifts the boson-fermion mass degeneracy that otherwise
allows the saturation of the Ward-Takahashi identities without the
need for a Goldstone particle.

Up to now, our analysis was restricted to chiral superfields and (in
the vacuum) unbroken supersymmetry. In order to come closer to
phenomenologically attractive models, it is necessary to understand
the behaviour of gauge fields~\cite{kk} as well as the interplay with
other mechanisms of supersymmetry breaking.

%=========================Danksagung===================================
\begin{ack}
  I would like to thank Prof.~Wilfried~Buchm\"uller for his continuous
  guidance in this project. The opportunity to finish this work under
  financial support by DESY is gratefully acknowledged.
\end{ack}

%===========================Literaturverzeichnis=======================

\end{fmffile}
\end{document}